\documentclass[aps,prl,showpacs,floatfix,twocolumn,byrevtex,superscriptaddress,showkeys]{revtex4-1}
\usepackage{amsmath}
\usepackage{amsthm}
\usepackage{amssymb}
\usepackage{amsmath}

\usepackage[english]{babel}
\usepackage{graphicx}
\usepackage{bm}
\usepackage{multirow}
\usepackage{dcolumn}
\usepackage{color}
\usepackage{textgreek}
\usepackage{hyperref}
\usepackage{todonotes}
\usepackage{verbatim}
\usepackage{soul}

\reversemarginpar

\begin{document}

\title{Weyl phonons in chiral crystals}

\author{Tiantian Zhang}\email{ttzhang@itp.ac.cn}
\affiliation{CAS Key Laboratory of Theoretical Physics, Institute of Theoretical Physics, Chinese Academy of Sciences, Beijing 100190, China}

\author{Zhiheng Huang}\email{This author contribute equally to this work.}
\affiliation{Beijing National Laboratory for Condensed Matter Physics, Institute of Physics, Chinese Academy of Sciences, Beijing 100190, China}
\affiliation{University of Chinese Academy of Sciences, Beijing 100049, China}

\author{Zitian Pan}
\affiliation{Beijing National Laboratory for Condensed Matter Physics, Institute of Physics, Chinese Academy of Sciences, Beijing 100190, China}
\affiliation{University of Chinese Academy of Sciences, Beijing 100049, China}

\author{Luojun Du}\email{luojun.du@iphy.ac.cn}
	\affiliation{Beijing National Laboratory for Condensed Matter Physics, Institute of Physics, Chinese Academy of Sciences, Beijing 100190, China}

\author{Guangyu Zhang}
	\affiliation{Beijing National Laboratory for Condensed Matter Physics, Institute of Physics, Chinese Academy of Sciences, Beijing 100190, China}
	\affiliation{Songshan Lake Materials Laboratory, Dongguan 523808, China}

\author{Shuichi Murakami}
\affiliation{Department of Physics, Tokyo Institute of Technology, Okayama, Meguro-ku, Tokyo 152-8551, Japan}

\begin{abstract}


\paragraph{\textbf{Abstract}}
Chirality is an indispensable concept that pervades fundamental science and nature, manifesting itself in diverse forms, $e.g.$, quasiparticles and crystal structures. Of particular interest are Weyl phonons carrying specific Chern numbers and chiral phonons doing circular motions. Up to now, they have been studied independently and the interpretations of chirality seem to be different in these two concepts, impeding our understanding. Here, we demonstrate that they are entangled in chiral crystals. 
Employing a typical chiral crystal of elementary tellurium (Te) as a case study, we expound on the intrinsic relationship between Chern number of Weyl phonons and pseudo-angular momentum (PAM, $l_{ph}$) of chiral phonons. 
We propose Raman scattering as a new technique to demonstrate the existence of Weyl phonons in Te, by detecting the chirality-induced energy splitting between the two constituent chiral phonon branches for Weyl phonons. 
Moreover, we also observe the obstructed phonon surface states for the first time. 


\end{abstract}


\maketitle

The breaking of continuous space-time symmetries in condensed matter solids has led to the emergence of novel quasiparticle excitations beyond the conventional Weyl-Dirac-Majorana classification.
Among them, Weyl quasiparticles with definite chirality have garnered significant attention. For example, unconventional Weyl fermions with various band degeneracies and topological charges have been uncovered~\cite{kruthoff2017topological,DWP_NEWFERMIONS,DWP_DOUBLE_HUANG,double_benjamin,PhysRevX.6.031003,double_bouhon,PhysRevLett.119.206402,chang2017unconventional,rao2019observation,takane2019observation,schroter2019chiral,lv2019observation,SciAd_C4_exp,miao2018observation,stenull2016topological,Lu622,DWP_MULTI_INPHOTONIC}, strengthening our understanding of elementary particles.
In addition to Weyl fermions, recent advances have revealed topological elementary excitations for bosonic systems, such as Weyl phonons in the THz energy range ($\sim$10 meV)~\cite{zhang2018double,li2018coexistent,xia2019symmetry,zhang2020twofold,liu2019ideal,liu2020symmetry,ding2022charge}. 
These excitations have drawn significant attention both theoretically and experimentally due to their potential to trigger novel quantum physical phenomena and applications, such as phonon Hall effects~\cite{zhang2010topological,qin2012berry,tang2021topological,im2022ferroelectricity} and topological thermal devices~\cite{liu2017model}. 
However, unlike Weyl fermions, which can be detected through various approaches such as angle-resolved photoemission spectroscopy~\cite{lv2015experimental,xu2015discovery} and electronic transport measurements~\cite{hosur2012charge,hosur2013recent,zhang2016signatures,liu2019magnetic}, the experimental probe of Weyl phonons is challenging and has only been achieved by inelastic x-ray spectroscopy (IXS) on their bulk wave functions in conjunction with first-principle calculations~\cite{miao2018observation,li2020observation}. 

\begin{figure}
  \centering
  \includegraphics[width=0.45\textwidth]{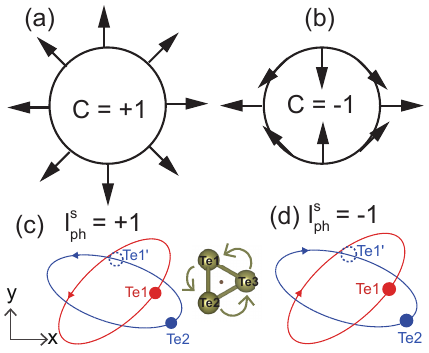}
  \caption{Spin-momentum locking configurations for Weyl points with (a) C = +1 and (b) C = $-1$.
  (c-d) Illustration for phonon modes with $l_{ph}^{s}$ = + 1 and $l_{ph}^{s}$ = $-1$, respectively. The insert figure shows the  top view of Te crystal with a right-hand structure. \label{fig0}}
\end{figure}

Apart from Weyl phonons, recent studies also uncover the concept of ``chiral phonon", characterized by negative circular polarization and pseudoangular momentum (PAM)~\cite{zhang2015chiral,zhang2022chiral,chen2021propagating,PhysRevB.103.L100409,tsunetsugu2023theory}. Chiral phonons can interact strongly with other circularly polarized excitations (e.g., photons) and thus be easily detected experimentally~\cite{malard2009raman,PhysRevB.92.024421,PhysRevB.97.174403,yang2018intrinsic,PhysRevB.97.195444,tatsumi2018interplay,drapcho2017apparent,chen2021probing, chen2019entanglement, li2019momentum, li2019emerging,zhu2018observation,ishito2023truly,ISHITO2023}. Up to now, Weyl and chiral phonons are studied independently, and the concept of ``chirality" seems to be different between these two concepts. The former is related to its (pseudo-)spin and momentum coupling patterns and is defined in reciprocal space, as shown in Figs.~\ref{fig0}(a-b), while the latter corresponds to phonon modes with atoms vibrating in a circular way in real space, as shown in Figs.~\ref{fig0}(c-d). If chiral phonons can entangle with Weyl phonons, they may provide the possibility to demonstrate the existence of Weyl phonons.

 In this study, we demonstrate that Weyl and chiral phonons are entangled in Te and further experimentally validate our theory by utilizing helicity-resolved Raman spectra. The evidence for the existence of Weyl phonons is revealed by directly tracking the chirality-induced energy splitting between the two constituent chiral phonon branches. Additionally, obstructed surface phonon states have been uncovered by Raman spectra for the first time. All the results are in perfect agreement with first-principles calculations, establishing Raman spectroscopy as a fast, non-destructive, and high spatial resolution probe for exploring topological phonons in chiral crystal structures, as well as the obstructed surface modes.

The Hamiltonian in the vicinity of conventional Weyl points can be succinctly characterized by $H(\mathbf{k})\propto \mathbf{S} \cdot \mathbf{k}$, where $\mathbf{S}$ denotes the rotation generator for spin-$\frac{1}{2}$ particles. 
Two bands of Weyl points display spin components of $\mathbf{S}\cdot\hat{k}=\pm\frac{\hbar}{2}$, leading to an inward or outward hedgehog configuration on each equal-energy surface. 
These distinct spin-momentum-locking properties determine the topological invariant referred to as the Chern number, which assumes the values of C = $\pm1$ and serves to differentiate between those two cases, as shown in Figs.~\ref{fig0} (a) and (b), respectively. In addition, the sign of the Chern number corresponds to the chirality of the Weyl points.
The conventional Weyl point with C = 1 may transform into an unconventional one when additional crystalline symmetries exist, such as a double Weyl point with C = 2 under threefold (screw) rotation symmetry {\cite{PhysRevB.90.155316,fang2015topological,zhang2020twofold}}.

\begin{figure}
  \centering
  \includegraphics[width=0.5\textwidth]{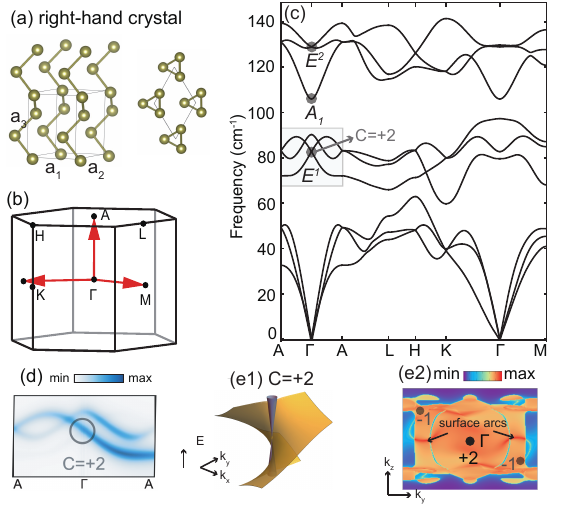}
  \caption{(a) Side (left panel) and top (right panel) views of elemental crystal tellurium with right-hand structure. (b-c) Brillouin zone and phonon band structure of tellurium. The twofold degenerate band marked by the light grey box at $\Gamma$ is a double Weyl phonon with C = +2. 
  (d) Dynamical structural factor simulation on the double Weyl phonon with C = +2 by following the $k$-path of $(4.0,5.0,-0.5)$-$(4.0,5.0,0.0)$-$(4.0,5.0,0.5)$. (e1) Illustration for the bulk dispersion (solid cone) and surface states (yellow sheets) for the double Weyl phonon at $\Gamma$ on the (010) surface. (e2) Surface arcs for the double Weyl phonon in tellurium.\label{fig1}}
\end{figure}

Tellurium crystals have two enantiomers, classified as the right-hand structure with space group $P3_{1}21$ or the left-hand structure with space group $P3_{2}21$. 
Figures~\ref{fig1} (a-c) depict the crystal structure, Brillouin zone (BZ), and phonon spectra for the right-hand tellurium crystal with space group, where the three Raman active branches are denoted by $E^1$, $A_1$, and $E^2$ with increasing frequency at $\Gamma$. According to first-principles calculations, the twofold phonon modes with $E^1$ and $E^2$ irreducible representations are both double Weyl phonons with $|$C$|$ = 2, protected by threefold screw rotation symmetry ($C_3'$) and time-reversal symmetry ($\mathcal{T}$)~\cite{DWP_fang,PhysRevB.96.045102,T_Nagaosa_2014,jian2015double}, and they are beyond the traditional classification. The low-energy effective $\mathbf{k}\cdot\mathbf{p}$ model for $E^1$ can be written as:
\begin{equation}
H(\mathbf{k})= \begin{pmatrix}
Ak_z & B(k_x-ik_y)^2\\
B(k_x+ik_y)^2 & -Ak_z
\end{pmatrix}
\end{equation}
where A and B are real constants. The Hamiltonian gives rise to a Weyl point with C = $+2$. 

\begin{figure*}
  \centering
  \includegraphics[width=0.8\textwidth]{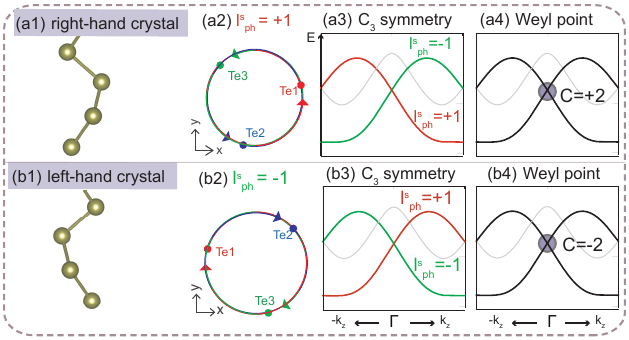}
  \caption{Entanglement between the chirality of crystal structure, Weyl phonon, and PAM. (a1) and (b1) are tellurium with right-hand and left-hand chiral crystal structures, respectively. 
  (a2) and (b2) are the atomic vibration for phonon modes with the same energy in different chiral crystal structures, showing opposite chiral motions in real space, and their PAMs shown in (a3) and (b3) are defined in terms of the threefold screw symmetry with respect to $k_z$ direction. 
  (a4) and (b4) are double Weyl phonons carrying opposite Chern numbers for (a1) and (b1), respectively. \label{fig2}}
\end{figure*}

 Figure~\ref{fig1} (d) illustrates the phonon dynamical structure factor (DSF) of Te simulated by first-principle calculations for the $E^1$ mode along the $k$-path of A$(4.0,5.0,-0.5)$-$\Gamma(4.0,5.0,0.0)$-A$(4.0,5.0,0.5)$. The nontrivial DSF intensity distribution arising from the topological bulk wave functions provides the possibility to verify the double Weyl phonon using IXS~\cite{miao2018observation,zhang2019phononic,li2020observation,jin2022chern}. In addition to the bulk topology property, Weyl phonons can also show topological surface states due to the bulk-surface correspondence (BSC). Figure~\ref{fig1} (e1) shows a double Weyl phonon with C = +2 (solid cone) associated with two helical surface states (yellow sheets), rotating in a right-hand way with increasing frequency and connecting to the bulk Weyl bands, sharing the same chirality with the Weyl point. Figure~\ref{fig1} (e2) displays the iso-energetic contour for the $E^1$ Weyl phonon along the (010) direction, where the double Weyl phonon is projected onto the center of the surface BZ, associated with two surface arcs. It is noteworthy that $E^1$ Weyl modes carry opposite Chern numbers between two enantiomers of chiral crystal tellurium, and the Chern number alters from +2 in the right-hand crystal to $-2$ in the left-hand crystal, as shown in Figs.~\ref{fig2} (a4) and (b4).

For 3D chiral systems with (screw) rotation symmetry, where both mirror and inversion symmetries are absent, the lattice chirality is closely related to the circular polarization and propagation direction of phonons~\cite{chen2022chiral}. 
Thus, in addition to describing topological properties, chirality can be also used to catalog chiral phonons with left/right-hand circular vibrating motions, carrying nonzero phonon circular polarization and a noninteger PAM $l_{ph}$. 
Since chiral phonons that we study in this paper are located in the vicinity of $\Gamma$, PAM $l_{ph}$ can be simplified as the intracell part $l_{ph}^{s}$, which describes the phase difference 
of rotational motions between two neighboring atoms under the (screw) rotation symmetry \cite{zhang2022chiral}. As the example shown in Fig.~\ref{fig0} (c), Te1$^{'}$ is rotated from Te1 by $C_{3}^{'}$ symmetry in the right-hand crystal, having a $\frac{2}{3}\pi$ phase advancing to Te2, and therefore this case corresponds to $l_{ph}^{s}$ = +1. 
Likewise, Fig.~\ref{fig0} (d) corresponds to $l_{ph}^{s}$ = $-1$. 

Figures \ref{fig2} (a2) and (b2) show the two-dimensional projections of atomic motions for the lower branches of $E^1$ modes along $\Gamma$-A path (i.e., $k_z$ direction) in right- and left-hand crystals, respectively. 
The lower branch of $E^1$ mode along $\Gamma$-A path in the right-hand (left-hand) crystal does counter-clockwise (clockwise) circular motions, which indicates the opposite $l_{ph}^{s}$ between right- and left-hand crystals. According to the three-fold rotation symmetry operation, $l_{ph}^{s}$ for the lower branches of $E^1$ modes along $\Gamma$-A path is +1 ($-1$) in right-hand (left-hand) crystal, as shown in Fig.~\ref{fig2} (a3) [Fig.~\ref{fig2} (b3)]. By comparing Fig.~\ref{fig2} (a4) and Fig.~\ref{fig2} (a3) [Fig.~\ref{fig2} (b4) and Fig.~\ref{fig2} (b3)], we conclude that Weyl phonons in Te are entangled with chiral phonons, as well as their chiral crystal structures, i.e., the Weyl phonon bands with C = +2 (C = $-2$) carries $l_{ph}^{s}$ = +1 ($-1$) for the lower phonon mode of $E^1$ along $\Gamma$-A path with the right(left)-hand crystal structure.

\begin{figure*}
  \centering
  \includegraphics[width=0.8\textwidth]{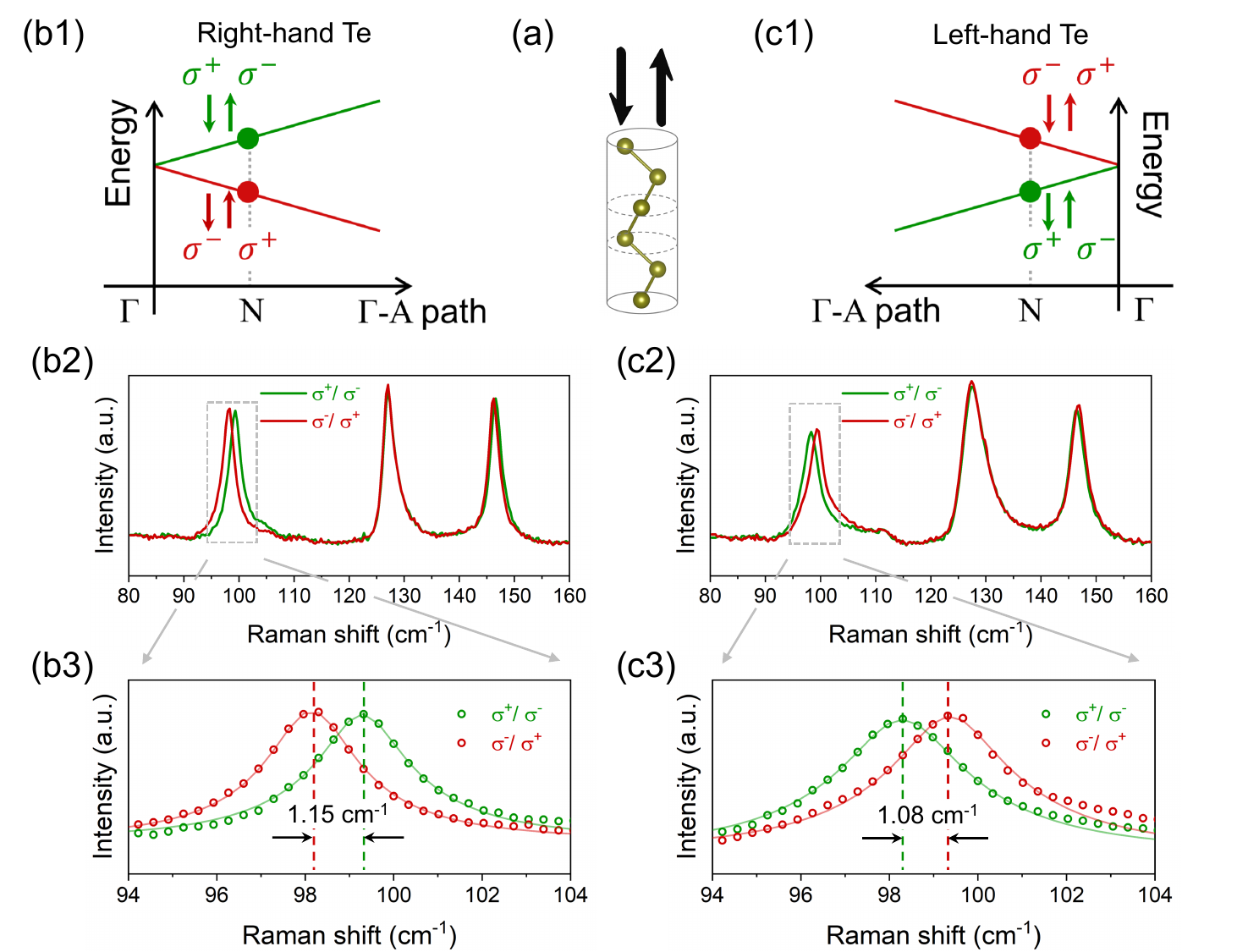}
  \caption{Weyl phonons detected by helicity-resolved Raman scattering. 
   (a) Backscattering geometry of helicity-resolved Raman spectroscopy on Te, where the incident/scattered photon propagates along the $\emph{c}$ axis. 
   (b1)/(c1) Schematic diagram of $E^1$ phonon dispersion along the $\Gamma-$A path, which is parallel to the $\emph{c}$ axis. The red (green) branch corresponds to chiral phonon mode with $l_{ph}^{s}$ = +1 ($-1$), which can be selectively detected under incident/scattered polarization configuration of $\sigma^{+}$/$\sigma^{-}$ ($\sigma^{-}$/$\sigma^{+}$) by backscattering Raman spectroscopy. $\sigma^{+}$ ($\sigma^{-}$) denotes left (right) circularly polarized light and propagates along the screw rotation axis. (b2)-(c2) Helicity-resolved Raman spectra in the $\sigma^{+}$/$\sigma^{-}$ (green) and $\sigma^{-}$/$\sigma^{+}$ (red) polarization conﬁgurations at 10 K for 633 nm excitation. (b3)-(c3) Zoom-in Raman spectra of $E^1$ mode, which clearly show the chirality-induced band splitting for the Weyl phonon. (b1)-(b3) are the results for Te with a right-hand crystal structure, while (c1)-(c3) are the results for Te with a left-hand crystal structure. 
  \label{fig3}}
\end{figure*}


As discussed above, Weyl phonons in chiral crystal Te are entangled with two chiral phonon branches holding opposite $l^{s}_{ph}$. This entanglement enables the detection of Weyl phonons through the fast, non-destructive, and high spatial resolution probe of helicity-resolved Raman scattering along the screw rotation axis (i.e., $\emph{c}$  axis), as illustrated in Fig.~\ref{fig3} (a). Typically, three conservation rules on the PAM, the pseudo-momentum, and the energy should be satisfied in the Raman scattering: (i) $l_i\ -$ $l_s$ = $l_{ph}^{s}$ $modulo$ 3, where $l_i$ ($l_s$) is psedoangular momentum for the incident (scattered) photons, and modulo three can be understood by considering the threefold screw rotation symmetry of Te and the Umklapp process~\cite{ishito2023truly,du2019lattice}; 
(ii) The momentum $N$ for phonons detected by backscattering Raman spectroscopy should arise from the wavenumber of photons \cite{PhysRevB.4.356}; (iii) Raman shift equal to the energy of the observed phonon modes at momentum $N$ shown in Figs.~\ref{fig3} (b1) and (c1). 
Here, with the wavelength $\lambda=633$ nm for the incident light, the corresponding wavenumber of momentum $N$ is $k=\frac{4\pi|n+i\eta|}{\lambda}\simeq0.8\times10^6$ cm$^{-1}$, which is about $\frac{1}{120}$ of the Brillouin zone along the $\emph{c}$ axis direction. 
$n = 2.3$ and $\eta = 3.5$ are the real and imaginary parts of the complex refractive index of Te, respectively \cite{PhysRevB.4.356}. Since the nonzero momentum leads to the energy splitting of the Weyl phonon carrying opposite chirality, Weyl phonons can be diagnosed by Raman spectroscopy.

As depicted in Figure~\ref{fig2} (a3), $l^{s}_{ph}$ is equal to $\pm1$ for the lower and upper branches of $E^1$ in Te with a right-hand crystal structure, respectively. 
In light of conservation rules, the upper (lower) branch can be selectivity detected under incident/scattered polarization configurations of $\sigma^{+}$/$\sigma^{-}$ ($\sigma^{-}$/$\sigma^{+}$), as shown in Fig.~\ref{fig3} (b1).  
Figure~\ref{fig3} (b2) is the polarized Raman spectra of right-hand Te, observed by using cross-circularly polarized detection with different circularly polarized light at 10 K, recorded with 633 nm excitation. 
The spectra exhibit two $E$ phonon modes ($E^1$: $\sim99$ cm$^{-1}$ and $E^2$: $\sim146$ cm$^{-1}$) and one $A_1$ mode ($\sim127$ cm$^{-1}$), which are in good agreement with first-principles calculations and previous results \cite{1970SSCom...8.1899T,PhysRevB.4.356}. Notably, as a nonchiral phonon mode, $A_1$ peaks appear almost identical in both configurations, while $E^1$ and $E^2$ phonon modes show significant energy splitting. 
Close-ups of Raman spectra for $E^1$ phonon mode (hollow circles) are shown in Fig.~\ref{fig3} (b3), fitted by the Voigt function (solid lines). 
The chirality-induced band splitting extracted from fitting is $\sim1.15$ cm$^{-1}$ for $E^1$ phonon mode, demonstrating the existence of two chiral phonons carrying opposite $l_{ph}^{s}$ and showing the existence of Weyl phonons in Te.

Figure~\ref{fig3} (c2) is the polarized Raman spectra of left-hand Te, which are in close resemblance to those of right-hand Te but exhibit opposing energy splits between $\sigma^{+}$/$\sigma^{-}$ and $\sigma^{-}$/$\sigma^{+}$ configurations. 
These results give information of $l_{ph}^{s}$ for the two chiral phonons and, consequently, the Chern number of the Weyl phonon in Te. Furthermore, both $l_{ph}^{s}$ and the Chern number of each band will change their sign if the chirality of the crystal is altered, showing good agreement with Figure~\ref{fig2}.


\begin{figure}
  \centering
  \includegraphics[width=0.48\textwidth]{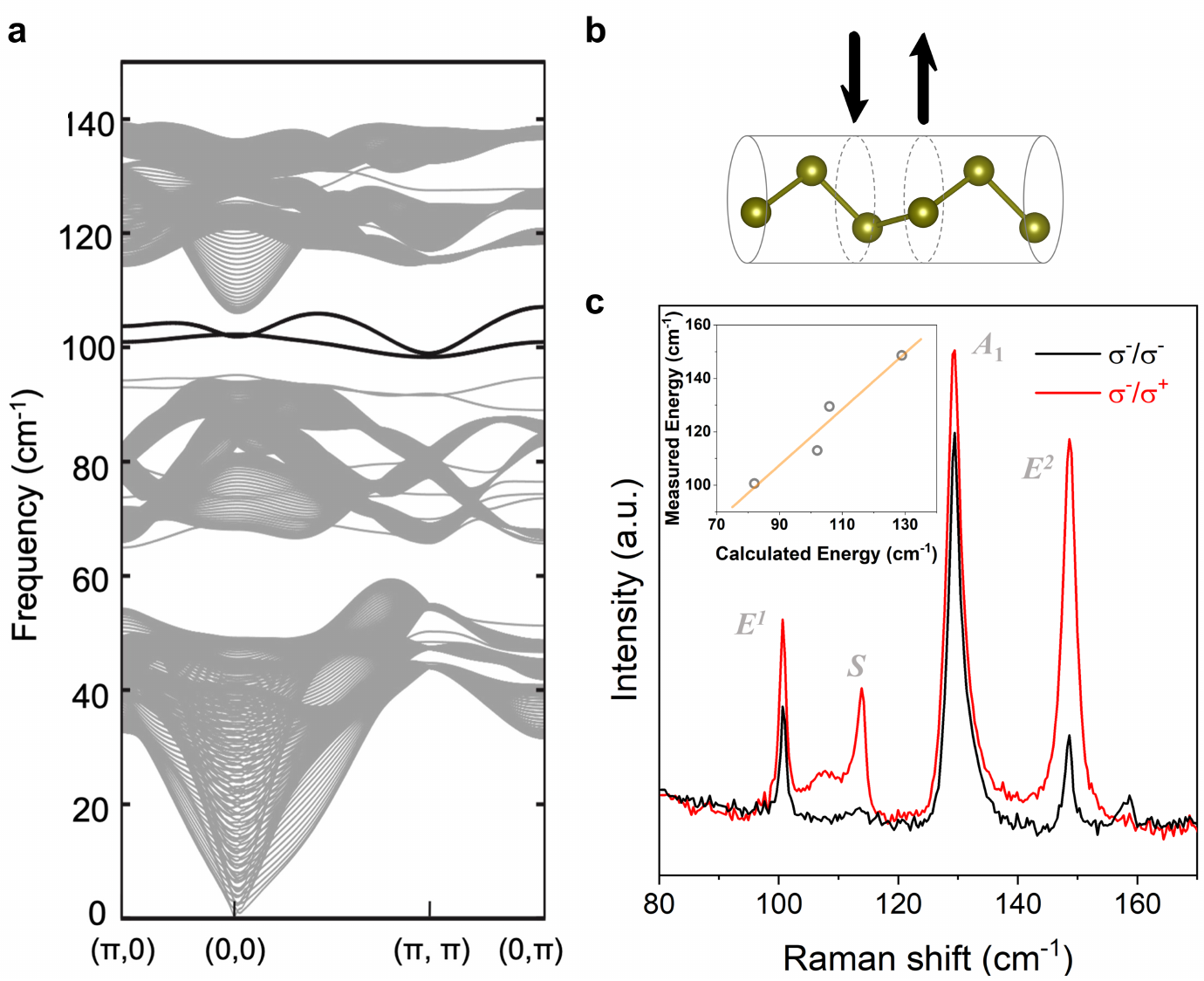}
  \caption{Demonstration for obstructed surface phonon modes. (a) First-principles calculations on right-hand Te with a slab configuration along [010] direction. Black solid lines at $\sim$100 cm$^{-1}$ are the obstructed surface phonon modes. (b) Backscattering geometry of helicity-resolved Raman spectroscopy on Te, where the incident and scattered photons are perpendicular to the $\emph{c}$ axis. (c) Co- (black) and cross- (red) circularly polarized Raman spectra under $\sigma^{-}$ excitation at 10 K. Inset: Fitting plot for the measured and calculated phonon energies, which shows a perfect match.
  \label{fig4}}
\end{figure}



The symmetry data (irreducible representation) for the lowest six phonon bands of Te is equal to that of an atomic insulator, so topological surface states are not expected in the frequency gap around 100 $cm^{-1}$. 
However, the symmetry data for the lowest six phonon bands is induced from the atomic positions where no real atoms reside, which makes Te belong to the recently proposed ``obstructed atomic insulator'', {which can be diagnosed by the real space topological invariant~\cite{po2018fragile,cano2018building,khalaf2021boundary,zhang2022large}. }
As a consequence, ``metallic'' surface states separate from the bulk bands will appear when the crystal is cleaved at these virtual sites on the surface.

Figure~\ref{fig4} (a) illustrates the surface phonon spectra obtained from first-principles calculations along the [010] direction, wherein the obstructed surface states marked by the black solid lines originate from the mismatch between the atomic position and the vibrating mass center. 
Since they spread over the whole surface Brillouin zone and are in the gap of bulk phonon spectra, they may therefore be observed by Raman scattering. 

Figure~\ref{fig4} (c) presents the co- (black) and cross- (red) circularly polarized Raman spectra under $\sigma{^-}$ excitation at 10 K, with the incident and scattered photons perpendicular to the $\emph{c}$ axis, as depicted in Figure~\ref{fig4} (b). 
Remarkably, a new phonon mode $S$ at $\sim115$ cm$^{-1}$ is detected under the cross-circularly polarized configuration ($\sigma^{-}$/$\sigma^{+}$) in addition to three bulk phonon modes.
The energies of the four experimentally observed modes are in excellent agreement with the first-principles calculated ones, as shown in the inset of Figure~\ref{fig4} (c), which confirms the $S$ mode being the obstructed surface phonon state. In the same setup,  when the surface is destroyed by a high-power laser, the $S$ mode will be efficiently quenched, which further proves that $S$ mode is a surface mode (See details in the Supplementary Material).

We highlight that, by combining Raman scattering and first-principle calculations, the existence of obstructed surface phonon modes is demonstrated for the first time. Note that surface phonon modes associated with Weyl phonons also spread widely in the surface Brillouin. However, they have the degeneracy in energy at $\Gamma$, which can not be distinguished by the Raman scattering.

%


The interplay between Weyl phonons and chiral phonons has received limited attention due to their independent investigations in different systems. 
In this study, we propose that the Weyl phonons carrying {nonzero} Chern numbers{, including $\pm1$, $\pm2$, $et\ al.$, } are intricately entangled with chiral phonons doing circular motions, with specific propagation directions and PAM{, in systems with $C_{3/4/6}$ (screw) rotation symmetries}. We present a novel experimental approach to probing and characterizing the chirality of Weyl phonons in chiral crystals that possess (screw) rotation symmetry, i.e., circularly polarized Raman scattering. Our approach is testified in chiral crystal Te, combined with first-principles calculations. Furthermore, we report the observation of obstructed phonon surface states for the first time by using Raman scattering and first-principles calculations, providing new insights into the phonon physics of chiral materials.

%
%
%
%

\section{\textbf{Acknowledge}}
This work is supported by the Strategic Priority Research Program of Chinese Academy of Sciences (CAS) under the grant No. XDB30000000, the National Natural Science Foundation of China (Grant No. 12047503), and Japan Society for the Promotion of Science (JSPS), KAKENHI Grant No. JP21K13865, No. JP22K18687 and JP22H00108, MEXT X-NICS under Grant No. JPJ011438, the National Key R\&D Program of China under grant No. 2021YFA1202900, the National Science Foundation of China (Grant Nos. 12274447, 12074412, 61888102, and 11834017).

\section{\textbf{References}}
\bibliography{reference.bib}
\bibliographystyle{unsrt}

\end{document}